\documentclass[%
aps,prl,twocolumn,10pt,amsmath,amssymb,nofootinbib,superscriptaddress,floatfix%
,nolongbibliography]{revtex4-2}

\usepackage[utf8]{inputenc}
\usepackage[usenames]{color}
\usepackage[dvipsnames]{xcolor}
\usepackage{graphicx}
\usepackage{amsmath}
\usepackage{amsfonts}
\usepackage{amssymb}
\usepackage{mathrsfs}
\usepackage{bm}
\usepackage{verbatim}
\usepackage{cancel}
\usepackage{comment}
\usepackage[normalem]{ulem}
\usepackage{CJK}
\usepackage{float}
\usepackage{textcomp}
\usepackage{booktabs}
\usepackage{subfigure}
\usepackage{isotope}
\usepackage{multirow,makecell}
\usepackage{braket}
\usepackage{hyperref}
\usepackage{environ}
\usepackage{suffix}
\usepackage{xspace}

\hypersetup{
  colorlinks=true,
  linkcolor=ForestGreen,
}

\graphicspath{{./figs/}}

\renewcommand{\vec}[1]{\mathbf{#1}}
\newcommand{\svec}[1]{\boldsymbol{#1}}

\newcommand{\isovec}[1]{\bm{#1}}

\newcommand{\chn}[3]{{{}^{#1}\!{#2}_{#3}}}
\newcommand{\cs}[2]{\chn{#1}{S}{#2}}
\newcommand{\chp}[2]{\chn{#1}{P}{#2}}
\newcommand{\cd}[2]{\chn{#1}{D}{#2}}
\newcommand{\cf}[2]{\chn{#1}{F}{#2}}
\newcommand{\cg}[2]{{}^{#1}{G}_{#2}}
\newcommand{\csd}{{\cs{3}{1}\text{-}\cd{3}{1}}}
\newcommand{\cpf}{{\chp{3}{2}\text{-}\cf{3}{2}}}
\newcommand{\cdg}{{\cd{3}{3}\text{-}\cg{3}{3}}}

\newcommand{\NNLO}{N$^2$LO}
\newcommand{\NNNLO}{N$^3$LO}

\newcommand{\ee}{\mathrm{e}}
\newcommand{\dd}{\mathrm{d}}
\newcommand{\vdelta}{\delta^{(3)}}

\newcommand{\mathspace}{\ \ }
\newcommand{\mathtext}[1]{\mathspace\text{#1}\mathspace}

\newcommand{\abs}[1]{|#1|}
\newcommand{\mean}[1]{\langle #1\rangle}

\newcommand{\ThreeH}{\isotope[3]{H}}

\newcommand{\FourHe}{\isotope[4]{He}}

\NewEnviron{subalign}[1][]{%
\begin{subequations}\begin{align}
  \BODY
\end{align}\label{#1}\end{subequations}
}

\NewEnviron{spliteq}{%
\begin{equation}\begin{split}
  \BODY
\end{split}\end{equation}
}

\newcommand{\CG}[6]{\ensuremath{C^{#5#6}_{#1#2,#3#4}}}
\WithSuffix\newcommand\CG*[3]{\ensuremath{C^{#3}_{#1,#2}}}

\newcommand{\SixJ}[6]{\left\{\!\!%
\begin{array}{ccc}%
{#1} & {#2} & {#3}\\[0.2em]%
{#4} & {#5} & {#6}%
\end{array}%
\!\!\right\}%
}

\newcommand{\couple}[3]{\left({#1}{#2}\right)\!{#3}}

\newcommand{\eg}{\textit{e.g.}\xspace}

\begin{document}

\title{Simplification of chiral nuclear forces
near the unitarity limit
}

\author{Songlin Lyu (吕松林)}
\affiliation{College of Physics, Sichuan University, Chengdu, Sichuan 610065, China}
\affiliation{Dipartimento di Matematica e Fisica, Universit\`a degli Studi della Campania ``Luigi Vanvitelli'', viale Abramo Lincoln 5 - I-81100 Caserta, Italy}
\affiliation{Istituto Nazionale di Fisica Nucleare, \\
Complesso Universitario di Monte  S. Angelo, Via Cintia - I-80126 Napoli, Italy}

\author{Lin Zuo (左林)}
\affiliation{College of Physics, Sichuan University, Chengdu, Sichuan 610065, China}

\author{Rui Peng (彭锐)}
\affiliation{College of Physics, Sichuan University, Chengdu, Sichuan 610065, China}
\affiliation{School of Physics, and State Key Laboratory of Nuclear Physics and Technology, \\
Peking University, Beijing 100871, China}

\author{Sebastian K\"onig}
\affiliation{Department of Physics, North Carolina State University, Raleigh, NC 27695, USA}

\author{Bingwei Long (龙炳蔚)}
\email{bingwei@scu.edu.cn}
\affiliation{College of Physics, Sichuan University, Chengdu, Sichuan 610065, China}
\affiliation{Southern Center for Nuclear-Science Theory (SCNT), Institute of Modern Physics, Chinese Academy of Sciences, Huizhou 516000, Guangdong, China}

\begin{abstract}
 Modern theory approaches for describing atomic nuclei often make use of on an
 effective theory that constructs the interaction between nucleons
 systematically based on Quantum Chromodynamics (QCD), exploiting constraints
 arising from the approximate chiral symmetry of QCD.
 The tensor nuclear force produced by one-pion exchange is an important feature
 that arises naturally in this framework.
 In this work we show that, however, the tensor force is suppressed by the large
 nucleon-nucleon scattering lengths in combination with the smallness of the
 pion mass.
 Based on this observation, we propose a new scheme for a chiral nuclear force
 that is able to describe $NN$ phase shifts up to the center-of-mass momenta $k
 \simeq 300$ MeV while treating pion exchange as a perturbation.
 Our much simplified leading-order force provides a microscopic explanation for
 the recent success of various short-range nuclear forces from the perspective
 of chiral effective field theory, and it shares with those approaches an
 approximate Wigner SU(4) symmetry, as well as the closeness to the unitarity
 limit (infinite nucleon-nucleon scattering lengths), as guiding principles.
 Compared to previous approaches to perturbative-pion interactions, our force
 also adjusts the ordering of short-range contact interactions, by means of
 which we overcome convergence problems of the expansion that were previously
 assumed to severely limit its usefulness.
 We demonstrate the performance of our approach with numerical calculations of
 $NN$ scattering up to fourth order, in addition to studies of $3N$ and $4N$
 bound-state properties.
\end{abstract}

\date{\today}

\begin{CJK*}{UTF8}{gbsn}
\maketitle
\end{CJK*}

Chiral effective field theory (ChEFT) has been the foundation of some of the most
popular nuclear forces developed and used over the past two
decades~\cite{Entem:2003ft, Epelbaum:2004fk, Nogga:2005hy, Long:2011xw,
PavonValderrama:2011fcz, Lu:2021gsb, Epelbaum:2014sza, Reinert:2017usi,
Tews:2022yfb, Entem:2017gor, RodriguezEntem:2020jgp, Nosyk:2021pxb,
Saha:2022oep}.
These potentials are usually derived from high orders in the so-called Weinberg
power counting, leading to a large amount of fine detail in the interaction
that arises from pion-exchange diagrams accompanied by contact (regulated
zero-range) interactions.
The latter feature unknown parameters that need to be determined from experiment or by matching to Quantum Chromodynamics (QCD).

While this approach is still broadly considered as \emph{the} avenue to describe
atomic nuclei from first principles, the past few years
have brought a surge of evidence that
the nuclear
interaction may be much simpler than these ``chiral potentials''
suggest~\cite{Platter:2004zs, Konig:2016utl, Vanasse:2016umz, Konig:2016iny,
Lensky:2016djr, Bansal:2017pwn, Konig:2019xxk, Kievsky:2018xsl, Lu:2018bat,
Gattobigio:2019omi, Deltuva:2020aws, Kievsky:2021ghz, Shen:2021kqr,
Meissner:2023cvo}.
These studies are based on (or similar to) the so-called Pionless EFT
in the sense that they are guided by the observation
that the unnaturally large nucleon-nucleon ($NN$) $S$-wave scattering lengths
give rise to certain universal features.
In Pionless EFT, the pions are ``integrated out'' and the remaining short-range
nuclear forces are then described solely by contact interactions.
This theory, as well as ChEFT, have most recently been reviewed
in~\cite{Hammer:2019poc}.

The leading-order (LO) $NN$ contact forces in Pionless EFT act only in $S$ waves
and can be written as
\begin{align}
 V_2^{(0)}(\vec{p}', \vec{p})
 = C_{\text{s}} + C_{\text{t}} \, \svec{\sigma}_1\cdot\svec{\sigma}_2 \,,
 \label{eqn:PL2N0}
\end{align}
where $1$ and $2$ label the nucleons, $\svec{\sigma}$ are the Pauli matrices in
spin space, and $\vec{p}$ ($\vec{p}'$) denotes the initial (final) relative
momentum between a pair of nucleons.
Note that until we introduce a regularization scheme (see below), there is
no actual momentum dependence in these contact forces.
The total LO force also includes a three-nucleon ($3N$) contact interaction that
is independent of momenta, spin or isospin~\cite{Bedaque:1998kg,Bedaque:1999ve}:
\begin{align}
 V_3^{(0)}(\vec{p}',\vec{q}'; \vec{p}, \vec{q}) = h \,,
 \label{eqn:PL3N0}
\end{align}
where $\vec{q}$ ($\vec{q}\,'$) is the third nucleon's momentum relative to the
pair in the initial (final) state.
We apply a separable regulator to
the potentials,
\begin{align}
 V_2(\vec{p}', \vec{p})
  \to &\ \ee^{{-}\frac{p'^4}{\Lambda^4}}
  V_2(\vec{p}' \, \vec{p}) \ee^{-\frac{p^4}{\Lambda^4}} \,, \\
 V_3(\vec{p}',\vec{q}'; \vec{p},\vec{q})
  \to\ & \ee^{{-}\frac{{\Pi^\prime}^4}{\Lambda^4}}
  V_3(\vec{p}',\vec{q}'; \vec{p},\vec{q})
  \ee^{{-}\frac{{\Pi}^4}{\Lambda^4}} \,,
\end{align}
where $\Lambda$ is an ultraviolet cutoff and
$\Pi^2/m_N \equiv (p^2 + \frac{3}{4}q^2)/m_N$ the $3N$ total kinetic energy.

The probably most extreme simplification of the nuclear force was developed
in Refs.~\cite{Konig:2016utl, Konig:2016iny, Konig:2019xxk}, which found that
the properties of (at least) light nuclei can be described by a
\emph{perturbative} expansion around the so-called ``unitarity limit,'' where
the $NN$ $S$-wave scattering lengths are infinite.
This leads to a parameter-free two-nucleon force at LO, with a single
three-nucleon datum required to fix the $3N$ contact interaction -- and all
further details (such as the actual finite values of the scattering lengths)
enter only at higher orders in strict perturbation theory.

In this approach, there is also a significantly increased amount of symmetry at
LO because the parameter-free $2N$ interaction fully realizes Wigner's symmetry,
SU(4) symmetry of the nucleonic spin-isospin quartet~\cite{Wigner:1936dx}.
A perturbative expansion of Pionless EFT around the SU(4) limit was
explored in Refs.~\cite{Vanasse:2016umz, Lin:2022yaf, Lin:2024bor}, and for
chiral potentials it was recently demonstrated that SU(4) symmetry significantly
constrains beta decays among light nuclei~\cite{LiMuli:2025zro}.
Another approach, resembling Pionless EFT, but adding explicit finite ranges as
model parameters, characterizes the nuclear interactions with Gaussian
potentials in order to capture the universal features of low-energy nuclear
physics~\cite{Gattobigio:2019omi, Deltuva:2020aws, Kievsky:2021ghz,
Gattobigio:2023fmo}, achieving impressive success for few-nucleon systems and
even nuclear matter~\cite{Kievsky:2018xsl}.
Similarly, in the context of nuclear Lattice EFT it was found that using
smeared, SU(4)-symmetric $NN$ and $3N$ interactions, can produce
remarkably accurate results for various nuclear ground states up to mass number
50~\cite{Lu:2018bat}, and also for the spectrum of
\isotope[12]C~\cite{Shen:2021kqr} and the \isotope[4]{He} monopole
resonance~\cite{Meissner:2023cvo}.

In light of all these findings, none of which are based on ChEFT, it is
interesting to contemplate a reorganization of chiral nuclear forces in which an
extremely simplified LO interaction emerges and provides the basis for
\textit{ab initio} calculations of strongly-correlated nuclear many-body
systems, leaving all finer details to small, perturbative corrections.
Assuming a binding energy per nucleon $E/A \sim 8$ MeV, the average momentum in
a typical nucleus can be roughly estimated as $\sqrt{2m_N E/A} \simeq 120$ MeV,
only slightly smaller than the pion mass $m_\pi = 139$ MeV~\cite{Konig:2016utl}.
It is therefore an open question whether Pionless EFT alone can still accomplish
this goal beyond the lightest nuclei, but ChEFT, for which $120$ MeV clearly is
a low-momentum scale, naturally suggests itself to provide these perturbative
details.
Such an expansion was in fact suggested decades ago, in the form of the
so-called KSW scheme~\cite{Kaplan:1998tg, Kaplan:1998we}, including
pion-exchange contributions perturbatively on top of the Pionless LO.
However, this approach was largely abandoned due to evidence that it has poor
convergence properties~\cite{Cohen:1999ds, Fleming:1999ee}; see
Refs.~\cite{Beane:2008bt, Kaplan:2019znu} for more recent developments.

In the following we argue that with certain modifications of the power counting,
a perturbative-pion interaction (PPI) \emph{can} in fact produce good
convergence properties for (at least) few-nucleon observables, and is thus
worthwhile to pursue.
Reference~\cite{Tews:2022yfb} (Chapter 9 therein) raised the question whether
the unitarity limit or the chiral limit ($m_\pi \to 0$) are ultimately more
relevant for light nuclei and highlighted several important connections.
Recent work~\cite{Teng:2024exc,Griesshammer:2025els} conjectures that the former,
and more broadly SU(4) symmetry, wins the chiral symmetry in terms of importance.
As a consequence, Refs.~\cite{Teng:2024exc,Griesshammer:2025els} propose to 
\emph{demote} the tensor component in the $\csd$ channel of the $NN$ potential 
generated by one-pion exchange (OPE) to a higher order than where it naively 
enters.

Taking a different point of view, we show that 
both limits, unitarity and chiral, are important and that they
work together to moderate the tensor force in $\csd$.
As the second cornerstone of our approach we argue that a set of
momentum-dependent contact forces must be arranged to enter at the same order as
OPE, and they work further to weaken the notoriously strong attraction of the
tensor force.
This part of the analysis is not limited to the $NN$ $S$-wave channels.
This is a correlation revealed by renormalization-group analysis of OPE at
relatively high momenta.
We proceed in the following to present our argument in detail and to show
results for both $NN$ scattering as well as light nuclei to demonstrate the 
performance and potential of our novel approach.

The OPE potential has the following coordinate-space form:
\begin{equation}
    V_\pi(\vec{r}) = \frac{\alpha_\pi}{4m_N}
    \isovec{\tau}_1\isovec{\cdot}\isovec{\tau}_2
    \left[
      \hat{S}_{12} T(r) + \svec{\sigma}_1 \cdot \svec{\sigma}_2 Y(r)
    \right] \,,
\end{equation}
where $\alpha_\pi \equiv g_A^2 m_N/(16\pi f_\pi^2)$ and $\isovec{\tau}$ are the
Pauli matrices in isospin space.
The operator $\hat{S}_{12}$ and the tensor and Yukawa potentials $T(r)$ and
$Y(r)$ are defined as follows:
\begin{align}
    \hat{S}_{12} &\equiv 3(\svec{\sigma}_1
    \cdot \hat{r})(\svec{\sigma}_2 \cdot \hat{r})
    - \svec{\sigma}_1 \cdot \svec{\sigma}_2 \,, \\
    T(r) &\equiv \frac{\ee^{{-}m_\pi r}}{r}
    \left(\frac{3}{r^2} + \frac{3m_\pi}{r} + m_\pi^2 \right) \,,
    \label{eqn:DefTr} \\
    Y(r) &\equiv m_\pi^2 \frac{\ee^{{-}m_\pi r}}{r} \,.
\end{align}
$\hat{S}_{12}$ is
a spherical tensor of rank two; therefore
it mixes the $\cs{3}{1}$ and $\cd{3}{1}$ components of the deuteron and
contributes to the mixing angle $\epsilon_1$ in $NN$ scattering; this is the (in)famous tensor force.
The KSW scheme conjectures an expansion in $\alpha_\pi Q$, where $Q$ denotes the
size of typical momenta of processes under consideration.
However, this expansion becomes obstructed by the large values of
$\hat{S}_{12}$ between certain partial waves.
For instance, the tensor potential $T(r)$ is around 20 times stronger than
the Yukawa potential $Y(r)$ in $\cs{3}{1}$ at $r = m_\pi^{{-}1}$:
\begin{equation}
    \frac{
    \braket{
     \cs{3}{1} |
     \isovec{\tau}_1\isovec{\cdot}\isovec{\tau}_2 \hat{S}_{12} T(m_\pi^{{-}1}) |
     \cd{3}{1}}
    }{
    \braket{
     \cs{3}{1} |
     \isovec{\tau}_1\isovec{\cdot}\isovec{\tau}_2
      \vec{\sigma}_1 \cdot \vec{\sigma}_2 Y(m_\pi^{{-}1}) |
     \cs{3}{1}}
    } = 14\sqrt{2} \, .
\label{eqn:TYratio}
\end{equation}
In fact, the unsatisfactory convergence radius of the KSW scheme in $\csd$ and
$\chp{3}{0}$ was attributed to the strength of the OPE tensor
force~\cite{Fleming:1999ee, Kaplan:2019znu}, while in other partial waves its
convergence is satisfactory~\cite{Wu:2018lai, Kaplan:2019znu}.

\begin{figure}
\centering
\includegraphics[scale=0.6]{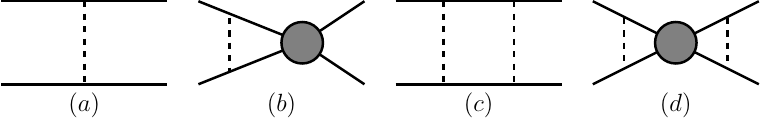}
\caption{Selected subleading Feynman diagrams of $NN$ scattering.
 The solid (dashed) represents the nucleon (pion).
}
\label{fig:NNFeyn}
\end{figure}

Despite the great strength of the tensor force, the mixing angle $\epsilon_1$, a prominent consequence of the tensor force, comes out
notably small ($\lesssim 2^\circ$ for center-of-mass momentum $k \leqslant 300$ MeV) from partial-wave analyses (PWA) of experimental data.
We find that this remarkable suppression is due to the simultaneous closeness of
nature to the unitarity limit in $\cs{3}{1}$
and to the chiral limit, by the following argument:
in our PPI scheme, the LO $\cs{3}{1}$ amplitude is generated by iterating
the LO contact potential~\eqref{eqn:PL2N0}, represented by the shaded blob in
Fig.~\ref{fig:NNFeyn}.
Due to the lack of tensor components, the mixing angle $\epsilon_1$ vanishes at
LO.
At NLO, the first-order perturbation in the tensor OPE contributes to $\epsilon_1$,
illustrated by Fig.~\ref{fig:NNFeyn}(a) and (b):
\begin{multline}
 \epsilon_{1,\pi} = \frac{m_N k}{4\pi} \frac{k}{(a^{-2}
 + k^2)^\frac{1}{2}} \\
 \null \times \Big[\frac{1}{a k}
  \braket{k, \cs{3}{1} | V_\pi | k, \cd{3}{1}}
 - \frac{g_A^2}{\sqrt{2}f_\pi^2}\,\Pi\Big(\frac{m_\pi}{k}\Big) \Big] \,,
 \label{eqn:VpiSD}
\end{multline}
where
\begin{equation}
 \Pi\Big(\frac{m_\pi}{k}\Big)
 \equiv \int_0^\infty \textrm{d} r' {r'}^2 n_0(k r') T(r') j_2(k r') \,,
\end{equation}
with $j_l(z)$ ($n_l(z)$) the spherical Bessel functions of the first (second)
kind.
Equation~\eqref{eqn:VpiSD} elucidates the impact of unitarity and chiral
limits: the first term inside the brackets indicates a suppression of the bare
OPE by $(ak)^{{-}1}$, vanishing in the unitarity limit, and the second term vanishes in the chiral
limit, where $T(r) \propto 1/r^3$, owing to
\begin{equation}
 \int_0^\infty \textrm{d}r r^2 n_0(k r) \frac{1}{r^3} j_2(kr) = 0 \,.
\label{eqn:N0Integral}
\end{equation}

This suppression of the tensor force by large $a$ and small
$m_\pi$ is remarkable, but it is limited to near on-shell kinematics, $p' \simeq p \simeq k$.
When OPE is iterated to the second order,
diagrammatically represented by Figs.~\ref{fig:NNFeyn}(c) and (d), the
contribution from $S$-wave intermediate states in sequences like $\cd{3}{1} \to
\cs{3}{1} \to \cd{3}{1}$ dominates the $\cd{3}{1}$ phase shifts, and this
is the primary culprit for the poor convergence of the KSW scheme in
$\cd{3}{1}$ (cf.\ Fig.~7 of Ref.~\cite{Fleming:1999ee}).
For guidance on how to mitigate this effect, we look to the
$\chp{3}{0}$ channel, where OPE alone is nonperturbative around $k \simeq
m_\pi$~\cite{Fleming:1999ee, Birse:2005umOPE, Wu:2018lai, Kaplan:2019znu}.
For a nonperturbative treatment of OPE, it was discovered early on that
renormalization of the scattering amplitude always requires a counterterm (i.e.,
a regulated contact interaction) at LO in partial waves where the tensor OPE is
attractive, including $\chp{3}{0}$~\cite{Beane:2000whLimitCycle,
Beane:2001bcTowards, PavonValderrama:2004nb,Nogga:2005hy}.
Analyzing the renormalization group (RG) flow of the $\chp{3}{0}$ counterterm 
$C_2^\chp{3}{0}$, Ref.~\cite{Peng:2020nyz} argued that
it can neutralize the singular attraction of the tensor force, so
much so that its sum with OPE forms the foundation for a new
perturbation theory -- which we adopt for PPI in $\chp{3}{0}$.

\begin{figure}
 \begin{center}
  \includegraphics[width=0.9\columnwidth]{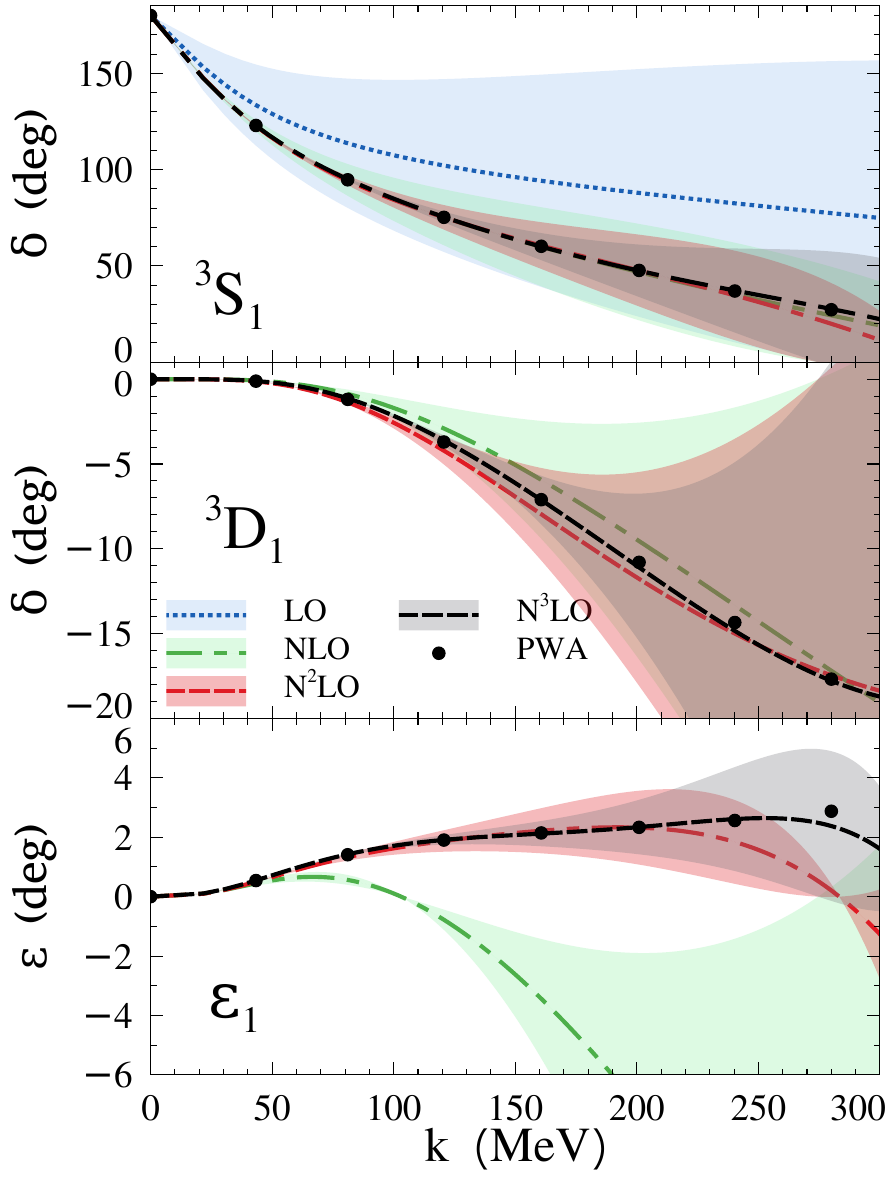}
 \end{center}
 \caption{$\csd$ phase shifts and mixing angles as a function of the
  center-of-mass momentum $k$ with cutoff value $\Lambda = 800$ MeV.
 The solid circles are the empirical phase shifts from the Nijmegen
 group~\cite{NNonline, Stoks:1993tb}.
 The green dot-dashed, red dashed, and dark solid lines correspond to NLO,
 N$^2$LO, and N$^3$LO respectively.
 The shaded bands indicate uncertainties estimated as $\pm (k \alpha_\pi)^{n-n_0+1}$
 relative to the central values at each order N$^n$LO, with $n_0$ the first non-vanishing
 order.
} \label{fig:3S1NNNLO}
\end{figure}

Following this idea, we now look for a counterterm to weaken the
tensor OPE in $\csd$ and conjecture that the lowest-dimension $SD$ mixing 
counterterm $C_2^{SD}$ can do just that, for it can counteract the tensor
force on the $S$-wave intermediate states in Fig.~\ref{fig:NNFeyn}(c) and (d).
Consequently, the NLO PPI potentials in $\csd$ are given by
\begin{align}
 \braket{\cs{3}{1},p' | V^{(1)}_{2} | \cs{3}{1},p}
 &= V_\pi + C_0^{(1)} + C_2^\cs{3}{1}({p}'^2 + p^2) \,, \\
 \braket{\cs{3}{1},p' | V^{(1)}_{2} | \cd{3}{1},p}
 &= V_\pi {-}C_2^{SD} p^2 \,,
\end{align}
where $C_0^{(1)}$ is the NLO correction to $C_0^{\cs{3}{1}}$.

At next-to-next-to leading order ({\NNLO}), one needs to consider second-order
perturbation theory of $V^{(1)}_{SD}$ on top of the LO $\cs{3}{1}$ amplitude.
Given the short-range nature of the LO $\cs{3}{1}$ potential, it is, incidentally,
much easier to identify the subleading counterterms needed to absorb
ultraviolet divergences at higher orders than in the nonperturbative-pion
case~\cite{PavonValderrama:2011fcz, Long:2011xw}.
More details about this and the PPI beyond {\NNLO} are provided in
the Supplemental Material.

In Fig.~\ref{fig:3S1NNNLO} we show the $\csd$ phase shifts and mixing angles
up to {\NNNLO} in PPI.
The agreement with the PWA at {\NNNLO} is excellent, with a discrepancy
$\lesssim 1^\circ$ up to $k = 300$ MeV.
Although the breakdown scale of a perturbative-pion EFT is expected to be
$\alpha_\pi^{-1} \simeq 270$ MeV~\cite{Kaplan:1998tg}, this performance
suggests that PPI may in fact enjoy a larger breakdown scale.
Overall (see other partial waves compiled in the Supplemental Material), we
find that PPI works almost as well as the nonperturbative-pion scheme
of Refs.~\cite{Long:2011xw, Long:2012ve, Wu:2018lai, Peng:2020nyz}.
This ``minimally modified Weinberg (MMW)'' scheme, as a chiral force without
$\Delta$ isobars, is expected to break down at $\delta \simeq 290$ MeV,
the $\Delta$-$N$ mass splitting, which is close to $\alpha_\pi^{-1}$.

\begin{figure}
 \centering \includegraphics[width=0.9\columnwidth]{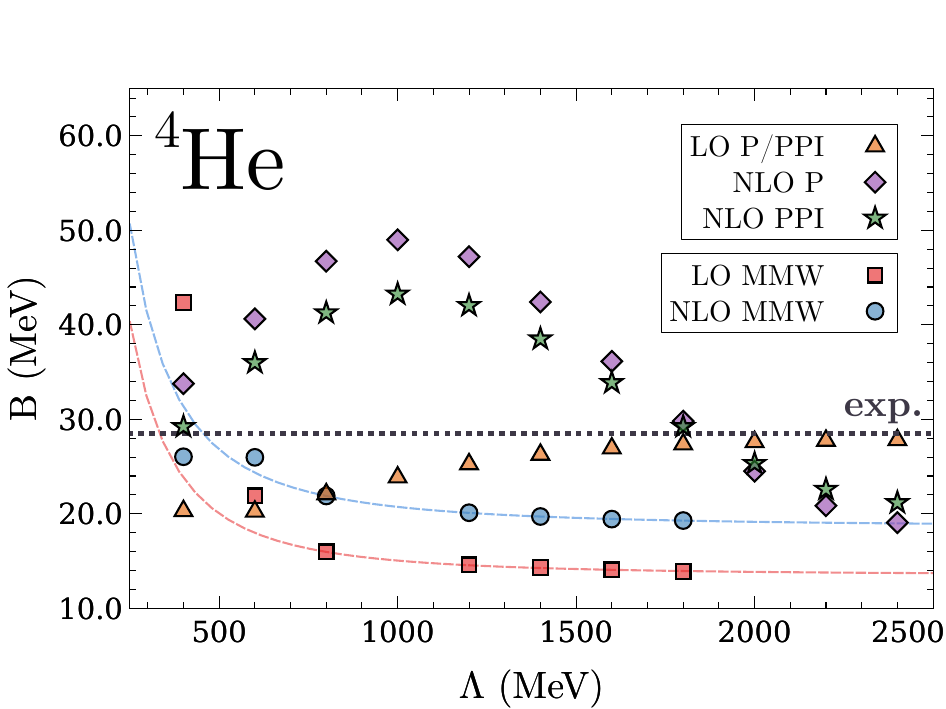}
 \caption{%
  Binding energy of $^4\text{He}$ as functions of $\Lambda$ with the Pionless,
  PPI, and MMW.
  For MMW, the dashes lines indicate fits of the form $B(\infty) + P(1/\Lambda)$,
  with $P$ a polynomial without constant term.
  \label{fig:PPIQ-He4}
 }
\end{figure}

To get a taste of how the PPI scheme performs in the few-nucleon sector, we
investigate the properties of $\isotope[3]{H}$ and $\isotope[4]{He}$ up to NLO.
$V_2^{(0)}$ and $V_3^{(0)}$ are spatially isotropic and thus conserve the
orbital angular momentum $L$, yet another LO symmetry of PPI.
Consequently, the tensor force does not correct the binding energy at NLO
for either \isotope[3]{H} or \isotope[4]{He} because their ground states have $L
= 0$.
Other parts of OPE still contribute at NLO, so a first objective is to
determine whether the
contact $4N$ force that appears at NLO in Pionless EFT~\cite{Bazak:2018qnu}
remains at that order in the PPI scheme.
Following Ref.~\cite{Bazak:2018qnu}, we examine in Fig.~\ref{fig:PPIQ-He4}
the cutoff dependence of the $\isotope[4]{He}$ binding energy calculated with
PPI.
The Coulomb interaction is omitted because it is known to be quite
weak for a tightly bound nucleus (\eg, although not 
model independent, \cite{Lu:2018bat} estimates a Coulomb correction 
$\lesssim 1$ MeV for \isotope[4]{He}), and in any case not relevant for the point
we study here.
Plotting the $\isotope[4]{He}$ binding energy against the EFT cutoff $\Lambda$,
we observe large variations that clearly indicate that both Pionless EFT and PPI 
need a contact $4N$ force at NLO for renormalization.
Pionless EFT and MMW results are included for comparison in Fig.~\ref{fig:PPIQ-He4}.

\begin{figure}
 \centering
 \includegraphics[width=0.9\columnwidth]{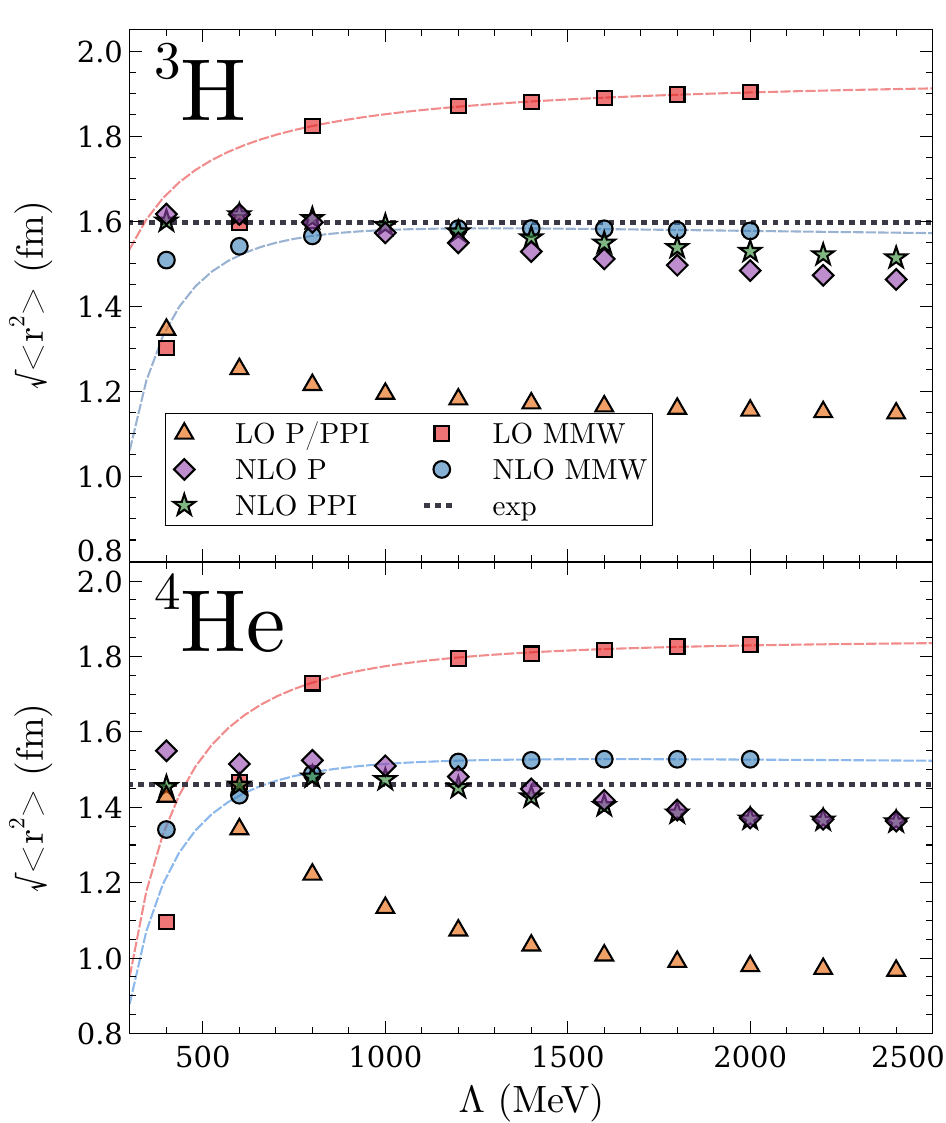}
 \caption{%
  The point charge radii of $\isotope[3]{H}$ (upper panel) and $\isotope[4]{He}$
  (lower panel) at LO/NLO from the Pionless, PPI, and nonperturbative-pion (MMW)
  scheme.
  The experimental point charge radii (dotted lines) are determined by
  converting the charge radii from Ref.~\cite{Angeli:2013xyz} according to
  Ref.~\cite{Friar:1975aa}.
  \label{fig:PPIQ-Rad}
 }
\end{figure}

We proceed to study radii at NLO, which can be corrected by the tensor OPE,
among other NLO forces, which modifies the wave functions.
In Fig.~\ref{fig:PPIQ-Rad} we show the point charge radii of
$\isotope[3]{H}$ and $\isotope[4]{He}$ as functions of $\Lambda$, with the NLO
$4N$ force included for $\isotope[4]{He}$ (fixed to reproduce the binding energy).
The radius calculations follow
Ref.~\cite{Konig:2019xxk}, with some updates summarized in the Supplemental
Material, and with other numerical improvements that will be
presented in a separate publication~\cite{Wu:2025xx}.
A recent perturbative Pionless calculation of these radii can be found in
Ref.~\cite{Mondal:2025wml}; our results (which neglect the Coulomb correction
in \isotope[4]{He} at NLO) are consistent with that determination.
Remarkably, we find that PPI provides only minuscule
improvements towards the experimental radii over Pionless EFT -- which are in fact
insignificant compared to the estimated uncertainty at NLO.
For these estimates, we assume low-energy scales
$Q_A = \sqrt{2 M_N B_A/A}$~\cite{Konig:2016utl}, with $B_3=B(\isotope[3]{H})$ and
$B_4=B(\isotope[4]{He})$, so $Q_3 \approx 73~\text{MeV}$ and 
$Q_4 \approx 115~\text{MeV}$.
Assuming $m_\pi \sim 139~\text{MeV}$ as Pionless EFT breakdown scale would suggest
a barely converging expansion for \isotope[4]{He}.
Our PPI 
gives $Q_4\alpha_\pi \simeq 0.43$ along with $Q_3\alpha_\pi \simeq 0.27$.
Although, as noted, the $NN$ results suggest that PPI may actually have a larger
breakdown scale, using these conservative estimates gives NLO radius predictions
$r_{\isotope[3]{H}} = 1.51\pm0.11$ fm and $r_{\isotope[4]{He}} = 1.36\pm0.25$ fm,
based on the calculated values at the largest cutoff shown in Fig.~\ref{fig:PPIQ-Rad}.
For Pionless EFT, the central values are similar, but the uncertainties are much larger.

Finally, we note that the PPI results converge at roughly the same rate as the MMW
ones, consistent with the similarity of the estimated respective breakdown scales:
$\alpha_\pi^{{-}1} \simeq \delta$.
Notably, however, the MMW radii tend to be larger, consistent with that
interaction's significant underbinding of the states at low orders (found for
\isotope[3]{H} already in Refs.~\cite{Nogga:2005hy,Song:2016ale} for a different
chiral power counting): in our calculations, we find 
$B(\isotope[3]{H}) \lesssim 5.3~\text{MeV}$ and
$B(\isotope[4]{He}) \lesssim 14~\text{MeV}$).
In PPI, both energies are input parameters that can be chosen arbitrarily close
to their experimental values.
Therefore, although less predictive, PPI provides an overall better description
of light nuclei than MMW.

In summary, we have identified a new perturbative-pion power counting in the
framework of ChEFT, with very promising performance and convergence properties.
At LO, PPI (like the abandoned KSW scheme) coincides with Pionless EFT and
therefore it is remarkably simple.
Higher orders systematically introduce the physics of pion-exchange in a controlled
way along with short-range contact operators.

Although we did not explicitly construct the LO force to reproduce exactly the
unitarity limit or a Wigner symmetric point, it is quite conceivable to do so,
as explored for Pionless EFT in
Refs.~\cite{Konig:2016utl,Vanasse:2016umz,Lin:2022yaf,Lin:2024bor}, and, as 
previously mentioned, in Refs.~\cite{Teng:2024exc,Griesshammer:2025els} in a
context similar to ours here.
While both the latter work and our approach are guided by similar
ideas, the conclusions are ultimately quite different.
Rather than imposing Wigner symmetry explicitly, we showed that it is an
emergent phenomenon.
Concretely, although the tensor force arising from OPE strongly breaks 
Wigner symmetry at NLO even if one assumes equal (and large) $NN$ $S$-wave 
scattering lengths at LO, we showed that the simultaneous closeness of 
nature to the chiral and unitarity limits suppress its impact on 
observables.
This is directly manifest in the smallness of $\epsilon_1$, which 
our Eq.~\eqref{eqn:VpiSD} explains up to NLO.
Beyond that, we found that promotion of the tensor contact operator to NLO 
further weakens the tensor OPE sufficiently to maintain the effect at 
higher orders.
With a similar argument mitigating the tensor force also in the 
$\chp{3}{0}$ channel, we arrive at the overall PPI scheme that provides
an excellent and convergent description of $NN$ scattering in all
relevant partial waves.

PPI is furthermore able to describe structural properties of light nuclei
with good precision and accuracy.
Notably, PPI gives almost the same results for the \isotope[3]{H} and 
\isotope[4]{He} radii as Pionless EFT.
At LO, this is by construction, but at NLO it is a nontrivial finding 
that may hint at why Pionless EFT performs better at describing light 
nuclei than one would naively expect.
Beyond the pure EFT paradigm, it also suggests an explanation for the
remarkable success of the very simple nuclear interactions used in 
Refs.~\cite{Gattobigio:2019omi, Deltuva:2020aws, Kievsky:2021ghz,
Gattobigio:2023fmo, Lu:2018bat, Shen:2021kqr, Meissner:2023cvo}, mentioned at
the outset as key motivation for our work.

Like Pionless EFT, PPI features a $3N$ force at LO and a $4N$ 
force at NLO, despite the explicit pion exchange entering at that order.
This provides indirect evidence that few-nucleon forces also play a larger
role in nonperturbative-pion
schemes, which has been suggested based on other arguments in
Ref.~\cite{Yang:2021vxa}.
We therefore conjecture that PPI will also prove useful for understanding
properties of many-nucleon systems, providing ample opportunity and inspiration for future investigations.

\acknowledgments
We thank Harald Grie{\ss}hammer, Matthias Heinz, Xincheng Lin, U.~van Kolck, and
Feng Wu for useful discussions and comments.
SK thanks the Institute for Nuclear Theory at the University of
Washington for its kind hospitality and stimulating research environment.
This research was supported in part by the INT's U.S. Department of Energy
grant No.~DE-FG02-00ER41132.
SK furthermore acknowledges the hospitality of the ECT*, where part of this
work was carried out.
This work was supported by the National Natural Science Foundation of China
(NSFC) under Grant Nos. 12275185 and 12335002 (BL).
The work of SK was supported in part by the U.S.\ National Science
Foundation (Grant No.~PHY--2044632).
This material is based upon work supported by the U.S.\ Department of Energy,
Office of Science, Office of Nuclear Physics, under the FRIB Theory Alliance,
Award No.~DE-SC0013617.
Computational resources for parts of this work were provided by the
high-performance computing cluster operated by North Carolina State University
and by the J{\"u}lich Supercomputing Centre.

\bibliography{PPIRefs.bib}

\clearpage
\section*{Supplemental Material}

\renewcommand{\section}{\subsection}

\setcounter{figure}{0}
\setcounter{table}{0}

\renewcommand{\thefigure}{SM\arabic{figure}}
\renewcommand{\thetable}{SM\Roman{table}}

\section{$NN$ Contact interactions}

The $NN$ contact interactions in Pionless EFT and ChEFT share the same functional
forms, which are most conveniently expressed in a partial-wave basis.
Up to {\NNNLO} in the PPI scheme, we need the following terms in $\cs{1}{0}$
and $\csd$:
\begin{align}
 &\braket{\cs{1}{0}, p'| V_\text{CT} | \cs{1}{0}, p}
 = C_0^{\cs{1}{0}} + C_2^{\cs{1}{0}} ({p'}^2 + p^2) \nonumber \\
 &\quad + C_4^{\cs{1}{0}} {p'}^2 p^2
 + C_6^{\cs{1}{0}} {p'}^2 p^2 ({p'}^2 + p^2)
 + \cdots \,,
 \label{eqn:VCT1S0}
\end{align}
\begin{widetext}
\begin{align}
 &\begin{pmatrix}
  \bra{\cs{3}{1},p'} \\
  \bra{\cd{3}{1},p'}
 \end{pmatrix}
 V_\text{CT}
 \begin{pmatrix}
  \bra{\cs{3}{1},p'} \\
  \bra{\cd{3}{1},p'}
 \end{pmatrix}^T
 =
 \begin{pmatrix}
  C_0^\cs{3}{1} & 0 \\
  0 & 0
 \end{pmatrix}
 + \begin{pmatrix}
  C_2^\cs{3}{1}({p}'^2 + p^2) & {-}C_2^{SD} p^2 \\
  {-}C_2^{SD} {p'}^2 & 0
 \end{pmatrix} \nonumber \\
 \null &\quad + \begin{pmatrix}
  C_4^\cs{3}{1} {p}'^2 p^2  & {-}C_4^{SD} p^2({p}'^2 + p^2) \\
  {-}C_4^{SD} {p'}^2 ({p}'^2 + p^2) & 0
 \end{pmatrix}
 + \begin{pmatrix}
  C_6^\cs{3}{1}{p}'^2 p^2({p}'^2 + p^2) & {-}C_6^{SD} {p}'^2 p^4 \\
  {-}C_6^{SD} {p}'^4 p^2 & C_4^{\cd{3}{1}} {p}'^2 p^2
 \end{pmatrix} + \cdots \,,
\label{eqn:VCT3SD1}
\end{align}
\end{widetext}
where the minus sign in front of the $SD$ mixing potentials $\sim C_{2n}^{SD}$
is adopted so that the value of $C_{2n}^{SD}$ is negative if the force it
represents is attractive.
Here $C_0^\cs{1}{0}$ and $C_0^\cs{3}{1}$ are related to $C_{\text{s}}$ and
$C_{\text{t}}$ by
\begin{align}
 C_0^\cs{1}{0} &= C_{\text{s}} - 3 C_{\text{t}} \,, \\
 C_0^\cs{3}{1} &= C_{\text{s}} + C_{\text{t}} \,.
\end{align}

In $P$ waves, the momentum dependence of the potentials is the same for all
possible values of the spin $s$ or angular momentum $j$:
\begin{multline}
 \braket{\chp{s}{j}, p'| V_\text{CT} | \chp{s}{j}, p}
 = C_2^{\chp{s}{j}} p' p \\
 \null + C_4^{\chp{s}{j}} p'p\, ({p'}^2 + p^2)
 + C_6^{\chp{s}{j}} p'p\,({p'}^2 p^2) + \cdots \,.
\label{eqn:VCT3P0}
\end{multline}
Except for $\cd{3}{1}$, $D$-wave or higher counterterms do not appear at
{\NNNLO} or lower orders.

\section{PPI power counting and phase shifts}

In the context of $NN$ scattering, nonperturbative treatment of a potential
means solving the Lippmann-Schwinger (LS) equation exactly for such a potential,
expressed in the operator form as
\begin{equation}
 T = V + T G_0 V \,,
\label{eqn:OptrLSeqn}
\end{equation}
where $T$ is the $T$ matrix of $NN$ elastic scattering, $V$ the regularized $NN$
potential, and $G_0$ the propagator of non-interacting $NN$ states.
The abstract operator equation can be written in the explicit momentum-space
partial-wave basis:
\begin{multline}
 T_{l' l}(p', p; E) =  V_{l' l}(p', p) \\
 \null + \frac{1}{2\pi^2} \sum_{l''} \int dq \, q^2\, V_{l' l''}(p', q)
 \frac{T_{l'' l}(q, p; E)}{E - \frac{q^2}{m_N} + i0} \,,
\label{eqn:PWLSeqn}
\end{multline}
where $l$, $l'$, and $l''$ are angular momenta and $E \equiv k^2/m_N$ the
center-of-mass energy.
To avoid confusion, we note that the following convention for partial-wave
states is used here:
\begin{equation}
 \braket{\vec{r} | p l m} = 4\pi i^l j_l(pr) Y_{lm}(\hat{r}) \,.
\end{equation}
The partial-wave $S$ matrix is related to the on-shell amplitude
$T_{l' l}(k, k; E)$ by
\begin{equation}
 S_{l' l} = \delta_{l' l} - i \frac{m_N k}{2\pi} i^{l' - l} T_{l'l} \,.
\end{equation}
The factor of $i^{l' - l}$ is a customary choice made by most partial-wave
analyses (see, e.g., Eq.~(17) in Ref.~\cite{Stoks:1993tb}).
While the exact history is difficult to pinpoint, we believe that this choice
has its origin in the coordinate-space LS equation commonly used in these
analyses, where the spherical Bessel function $j_l(kr)$, as opposed to $i^l
j_l(kr)$, is chosen as the incoming wave function.

In our notation, the LO potential $V^{(0)}$ is always nonperturbative regardless
of the counting scheme adopted, and thus the LO $NN$ partial-wave amplitude is
obtained by solving the following equation:
\begin{equation}
 T^{(0)} = V^{(0)} + T^{(0)} G_0 V^{(0)} \,.
\end{equation}
On top of the nonperturbative LO, the subleading amplitudes are generated by
a distorted-wave expansion.
For instance, the NLO amplitude $T^{(1)}$ is given by
\begin{multline}
 T^{(1)} = V^{(1)} + T^{(0)} G_0 V^{(1)} \\
 \null + V^{(1)} G_0 T^{(0)} + T^{(0)} G_0 V^{(1)} G_0 T^{(0)} \,.
\end{multline}
In Table~\ref{tab:PCPPI}, we tabulate the arrangement of contact interactions
in the PPI up to {\NNNLO}, grouped by the partial waves they act on.
The two-pion exchange (TPE) potential is the one with only chiral index $\nu =
0$ vertices, and its expression is taken from Ref.~\cite{Kaiser:1997mw}.
Except for $\csd$ and $\chp{3}{0}$, the PPI power counting follows that of the
KSW scheme~\cite{Kaplan:1998tg, Fleming:1999ee}.

Reference~\cite{Peng:2020nyz} worked out the PPI counting for $\chp{3}{0}$ by
pitching $C_2^\chp{3}{0}p' p$ against OPE at NLO.
At {\NNLO}, the relevant loop integral $I_{2,\chp{3}{0}}$ arising from
iterating $C_2^\chp{3}{0}p' p$ is
highly divergent:
\begin{equation}
 I_{2,\chp{3}{0}} \sim \frac{m_N}{2\pi^2}\, (C_2^\chp{3}{0})^2 p' p
 \int^\Lambda \textrm{d}l l^4 \frac{1}{k^2 - l^2} \,.
\end{equation}
As a result one needs $C_4^\chp{3}{0}$ to absorb the divergence at {\NNLO}, and
so on.
Power counting at {\NNLO} and {\NNNLO} in $\csd$ follows the same reasoning once
we decide to pair $C_2^{SD}$ with OPE at NLO, as argued in the main text.
In particular, $C_4^{\cs{3}{1}}$ appears at {\NNLO} for renormalization purpose. Reference~\cite{Fleming:1999ee} actually noted that $C_4^{\cs{3}{1}}$ at {\NNLO} would afford a much better description of the phase shifts, but could not find a theoretical argument for such a choice beyond a phenomenological observation.

\begin{table}
 \centering
 \begin{tabular}{c|cccc}
  & LO & NLO & {\NNLO} & {\NNNLO}\\
  \hline\hline
  $\pi$ & & OPE & & TPE \\
  \hline\hline
  $\cs{3}{1}$\, $\cs{1}{0}$ & $C_0$ \qquad & $C_2$ & $C_4$ & $C_6$\\
  \hline\hline
  $SD$ & & $C_2$ & $C_4$ & $C_6$  \\
  \hline\hline
  $\cd{3}{1}$ & & & & $C_4$ \\
  \hline\hline
  $\chp{3}{0}$ & & $C_2$ & $C_4$ & $C_6$ \\
  \hline\hline
  $\chp{3}{1, 2}$\, $\chp{1}{1}$ & &  & & $C_2$
 \end{tabular}
 \caption{%
  The PPI power counting of $NN$ forces.
  Rows labeled by a partial wave describes contact interactions acting in that
  partial wave.
  Pion exchanges are in the row led by ``$\pi$''.
  \label{tab:PCPPI}
 }
\end{table}

To determine the low-energy constants (LECs) displayed in
Table~\ref{tab:PCPPI}, we fit them numerically to reproduce the empirical phase
shifts from the Nijmegen partial-wave analysis (PWA)~\cite{NNonline,
Stoks:1993tb} up to $k = 300$ MeV.
We note that it is the $S$ and $P$ waves where tuning the LECs makes a
significant impact, whereas $\epsilon_2$, $\cd{1}{2}$, $\cd{3}{2}$, $\cdg$,
and $\cf{3}{2}$ do not involve contact LECs up to {\NNNLO}.
The values of pion-related physical constants are as follows: the axial coupling
constants $g_A = 1.29$, pion decay constant $f_\pi =
92.4$ MeV, average nucleon mass $m_N = 939$ MeV.

As explained in the main text, $C_2^{SD}$ is anticipated to balance the OPE
tensor force at {\NNLO} in the off-shell region; therefore, we constrain its
value by a combined fit to $\cd{3}{1}$ and $\epsilon_1$ of the PWA at {\NNLO}.
Doing so sacrifices perfect agreement with the PWA value for $\epsilon_1$ at
NLO, but the discrepancy is within the expected EFT uncertainty, 
and overall this choice improves
the convergence; LECs at yet higher orders will ultimately produce agreement
with the PWA.
In $\cs{1}{0}$ where the KSW scheme applies, it has been
argued~\cite{Fleming:1999ee, vanKolck:1998bw} that $C_4^\cs{1}{0}$ at {\NNLO}
does nothing more than absorbing the divergence, without carrying any
information about the shape parameter that enters at {\NNLO}.
We do not implement this in an exact manner because it is not straightforward to
do so in our fitting procedure.
At any rate, we do not expect this subtle difference in dealing with
$C_4^\cs{1}{0}$ to fundamentally change the convergence property of the PPI.
We have also used a fitting strategy in $\chp{3}{0}$ different from that of
Ref.~\cite{Peng:2020nyz}, in favor of a smaller incremental change from NLO to
{\NNLO}.

\begin{figure*}
 \centering
 \includegraphics[width=1.0\textwidth]{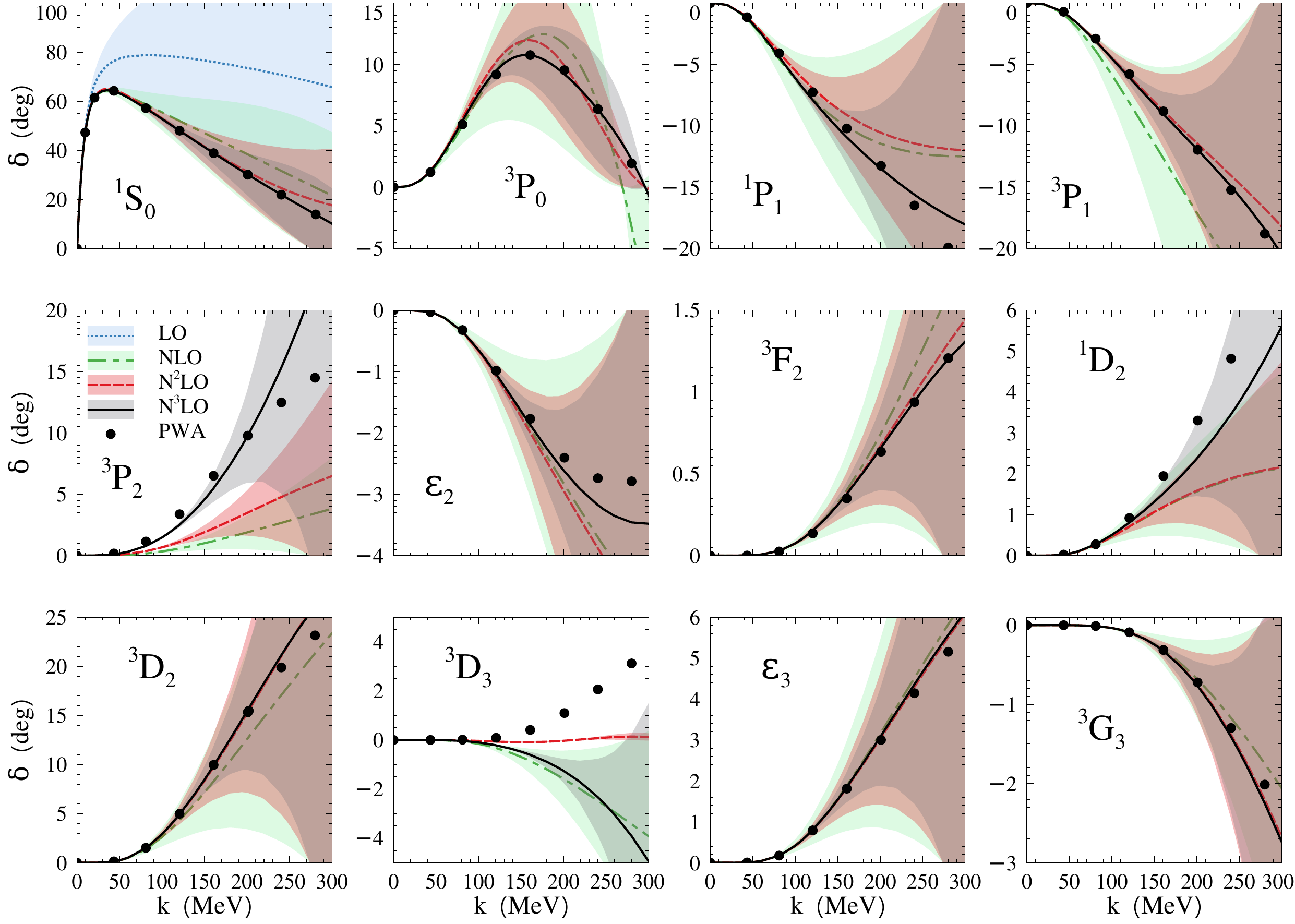}
 \caption{%
  $NN$ phase shifts and mixing angles from the PPI power counting, as functions
  of the center-of-mass momentum $k$ with cutoff value $\Lambda = 800$ MeV.
  The green dot-dashed, red dashed and black solid correspond to NLO, N$^2$LO
  and N$^3$LO respectively.
  The solid circles are the empirical phase shifts from the Nijmegen
  group~\cite{NNonline, Stoks:1993tb}.
  \label{fig:PPIPS}
 }
\end{figure*}

The phase shifts and mixing angle in $\csd$ are shown in the main text.
The comparison with the Nijmegen phase shifts for other channels is additionally
shown here in Fig.~\ref{fig:PPIPS}.
The convergence appears to be slow in $\cd{3}{3}$ and the disagreement with the
PWA is around four degrees at $k = 200$ MeV.
However, we note that Ref.~\cite{Kaplan:2019znu} showed that still higher-order
iterations of OPE will eventually converge in $\cd{3}{3}$.

\section{$NN$ forces in few-nucleon calculations}

When calculating the properties of $\isotope[3]{H}$ and $\isotope[4]{He}$ up to
NLO, we use $NN$ forces in channels with orbital angular momentum $l \leqslant
2$.
However, we retain the higher-wave components of the coupled channels to allow
the OPE tensor force to fully act: $\cs{1}{0}$, $\csd$, $\chp{1}{1}$,
$\chp{3}{0,1}$, $\cpf$, $\cd{1}{2}$, $\cd{3}{2}$, and $\cdg$.

The nonperturbative-pion MMW (``minimally modified Weinberg'') power counting
is based on Refs.~\cite{Nogga:2005hy, Long:2011xw, Long:2012ve, Wu:2018lai,
Kaplan:2019znu}, as compiled in Table~\ref{tab:PCMMW}.
In this counting, no three-nucleon forces appear until {\NNLO} in the version
without the delta-isobar degrees of freedom~\cite{Hammer:2019poc}.
The OPE potential is treated nonperturbatively in $\cs{1}{0}$, $\csd$, and
$\chp{3}{0}$, and in each of these three channels a counterterm is needed at LO.
In all other channels, OPE is treated perturbatively, thus appearing at NLO. In
$\cs{1}{0}$, a momentum dependent counterterm $C_2^\cs{1}{0}$ enters at NLO
while the NLO potentials in $\csd$ and $\chp{3}{0}$ vanish.

\begin{table}
\centering
 \begin{tabular}{c|cc}
  & LO & NLO \\
  \hline\hline
  $\cs{1}{0}$ & $C_0$ + OPE & $C_2$ \\
  \hline\hline
  $\csd$ & $C_0$ + OPE &    \\
  \hline\hline
  $\chp{3}{0}$ & $C_0$ + OPE &  \\
  \hline\hline
  all others & & OPE
 \end{tabular}
 \caption{
  The MMW power counting of $NN$ contact forces.
  \label{tab:PCMMW}
 }
\end{table}

\begin{figure*}
 \centering
 \includegraphics[width = 0.8 \textwidth]{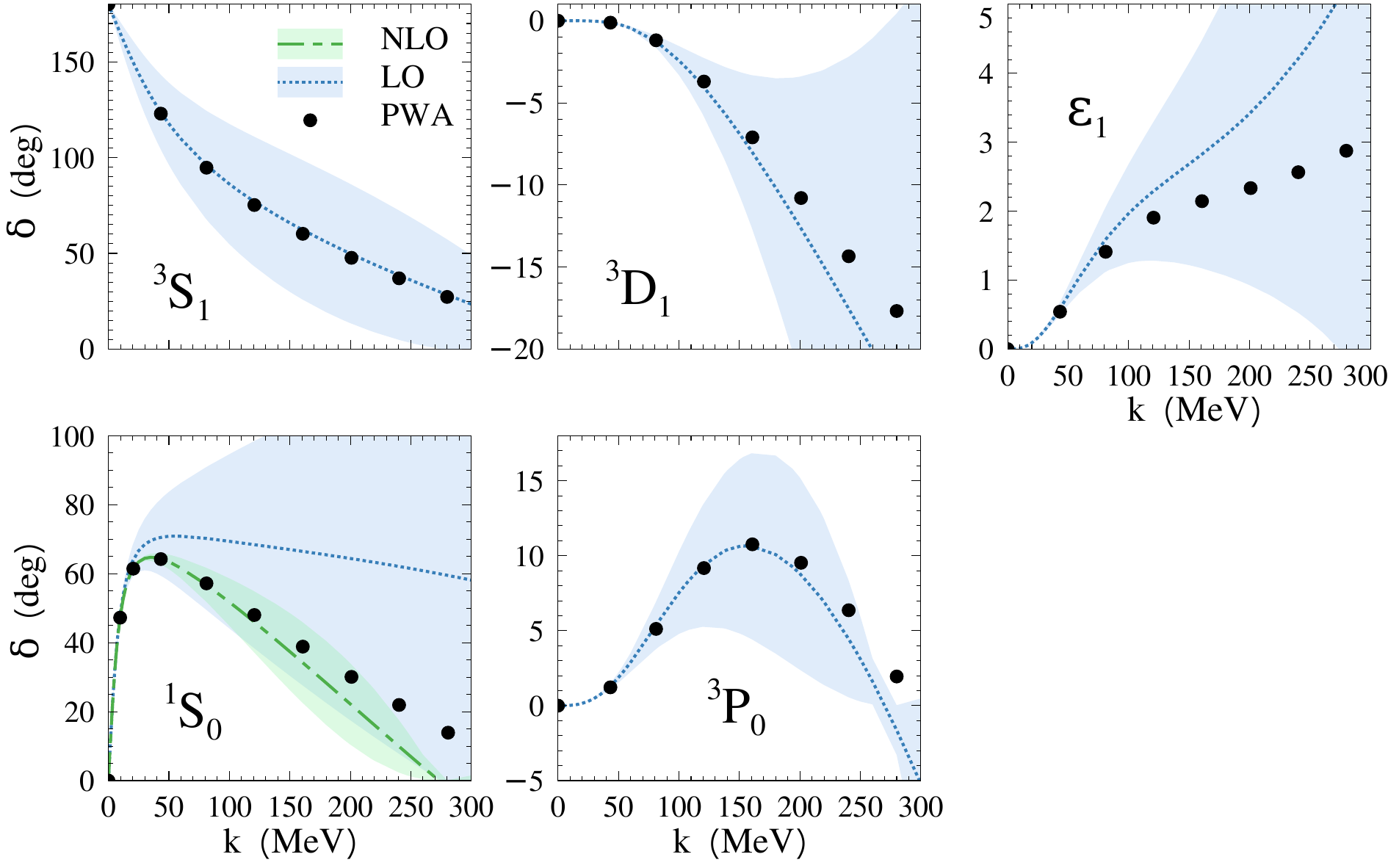}
 \caption{
  The $NN$ phase shifts with $MMW$ power counting up to NLO.
  For the explanation of the symbols, see Fig.~\ref{fig:PPIPS}.
  \label{fig:MMWPS}
 }
\end{figure*}

The $NN$ phase shifts from the MMW scheme are plotted up to NLO in
Fig.~\ref{fig:MMWPS}.
Only $\cs{1}{0}$, $\csd$, and $\chp{3}{0}$ are shown because there is only the
OPE potential acting in other channels at NLO, and this is already displayed as
the PPI NLO in Fig.~\ref{fig:PPIPS}.

\section{$3N$ and $4N$ calculations}

For the three- and four-nucleon observables discussed in the main text,
we utilize the Faddeev/Faddeev-Yakubowsky framework discussed in
Ref.~\cite{Konig:2019xxk}.
In this approach, we write a $3N$ or $4N$ state generically as
\begin{equation}
 \ket{\Psi} = \ket{\Psi^{(0)}} + \ket{\Psi^{(1)}} + \cdots \,,
\end{equation}
where the superscripts indicate the perturbative expansion induced by the
expansion of the nuclear interaction, as discussed in the main text.
At LO, $\ket{\Psi^{(0)}}$ is determined by solving the Faddeev
($3N$) or Faddeev-Yakubowsky ($4N$) equation, yielding both the binding energy as
well as the initial component from which the full state can be reconstructed.
The NLO shift for the energy is obtained by evaluating
the expectation value of the NLO potential corrections for the state
$\ket{\Psi^{(0)}}$.
Subsequently, the NLO state correction $\ket{\Psi^{(1)}}$ can be obtained by
solving inhomogeneous Faddeev/Faddeev-Yakubowsky equations, as described in
detail in Ref.~\cite{Konig:2019xxk}, along with the relevant potentials for
Pionless EFT at LO and NLO.

To calculate radii, we define the charge form factor $F_C$ as
\begin{equation}
 F_C(q^2) = \braket{\Psi^{}|\hat{\rho}(\vec{q})|\Psi} \,,
 \label{eq:FC}
\end{equation}
where $\hat{\rho}(\vec{q})$ is the relevant current operator, namely the charge
density for momentum transfer $\vec{q}$.
In principle, the current operator should be expanded perturbatively like all
other quantities, $\hat{\rho} = \hat{\rho}^{(0)} + \hat{\rho}^{(1)} + \cdots$,
but $\hat{\rho}^{(1)}$ vanishes and we do not need higher orders in this work.
Therefore, we only include the leading one-body charge density
$\hat{\rho}^{(0)}$, which can be written as
\begin{multline}
 \rho^{(0)}(\vec{q};\vec{p},\vec{p'})
 = \braket{\vec{p}|\rho^{(0)}(\vec{q})|\vec{p}'}
 = e\frac{1+\tau_3^{(i)}}{2} \\
 \null\times\vdelta\!\left(\vec{p}-\vec{p}'-\frac{\vec{q}}{2}\right) \,,
 \label{eq:rho-0}
\end{multline}
with $e$ the proton charge (which we set to unity in the following) and the
superscript on the isospin Pauli matrix $\tau_3$ indicating the nucleon that the
current is coupling to (assumed to be part of a pair with relative momentum
$\vec{p}$ ($\vec{p}'$) after (prior to) to the interaction with the photon.
The perturbative expansion of $F_C$,
\begin{equation}
 F_C(q^2) = F_C^{(0)}(q^2) +  F_C^{(1)}(q^2) + \cdots ,,
\end{equation}
follows then entirely from the expansion of the states $\ket{\Psi} =
\ket{\Psi^{(0)}} + \ket{\Psi^{(1)}} + \cdots$.
From the charge form factor we can extract point charge radii as
\begin{equation}
 \mean{r_0^2}
 = {-}\dfrac{1}{6}\dfrac{\dd}{\dd(q^2)} F_C(q^2)\Big|_{q^2=0}
 \mathtext{,}
 \mean{r_0} \equiv \sqrt{\mean{r_0^2}} \,,
\label{eq:r2-FC}
\end{equation}
the perturbative expansion of which follows from the expansion of $F_C(q^2)$,
as discussed in Ref.~\cite{Konig:2019xxk}.
Following Refs.~\cite{Miyagi:2015kza,Friar:1975aa}, the experimental values of
the point charge radii are defined as
\begin{equation}
 \mean{r_0^2}_{\ThreeH}
 = \mean{r^2}_{\ThreeH} - \mean{r^2}_p - 2\mean{r^2}_n
\end{equation}
for the \isotope[3]{H}, and as
\begin{equation}
 \mean{r_0^2}_{\FourHe}
 = \mean{r^2}_{\FourHe} - 2\mean{r^2}_p - \mean{r^2}_n
\end{equation}
for \isotope[4]{He}.\footnote{The expression for $\mean{r_0^2}_{\FourHe}$
 differs from what is stated in Ref.~\cite{Konig:2019xxk}, which had an
erroneous factor $2$ in front of $\mean{r^2}_n$.}
The very small Darwin-Foldy correction is neglected here.

The $3N$ and $4N$ wave functions are expressed in a Jacobi coordinate
partial-wave basis, with states written as $\ket{u_1 u_2;s}$ for the $3N$
system, and as $\ket{u_1 u_2 u_3;a}$ for the $4N$ system.
The Jacobi coordinates are defined as
\begin{subalign}[eqs:Jacobi-34]
 \vec{u}_1 &=
 \frac12(\vec{k}_1-\vec{k}_2) \,, \\
 \vec{u}_2 &=
 \frac23[\vec{k}_3-\frac12(\vec{k}_1+\vec{k}_2)] \,, \\
 \vec{u}_3 &=
 \frac34[\vec{k}_4-\frac13(\vec{k}_1+\vec{k}_2+\vec{k}_3)] \,,
\end{subalign}
where $\vec{k}_i$ labels the individual nucleon momenta.
The relevant collections of discrete quantum numbers (angular momentum,
spin, isospin; see Ref.~\cite{Konig:2019xxk} for details) are
\begin{equation}
 \ket{s} = \ket{%
  \couple{l_2}{\couple{\couple{l_1}{s_1}{j_1}}{\tfrac12}{s_2}}{J};
  \couple{t_1}{\tfrac12}{T}
 }
\label{eq:s}
\end{equation}
and
\begin{multline}
 \ket{a} = \ket{\!
   \couple{l_2}{\couple{\couple{l_1}{s_1}{j_1}}{\tfrac12}{s_2}}{j_2},
   \couple{l_3}{\tfrac12}{j_3},
   \couple{j_2}{j_3}{J}} \\
  \null \times \ket{
   \couple{\couple{t_1}{\tfrac12}{t_2}}{\tfrac12}{T}
 } \,.
 \label{eq:a}
\end{multline}
While in general the total charge operator for $A$ nucleons is given by
\begin{equation}
\hat{\rho}^{(0)} = \sum_{i=1}^A \hat{\rho}^{(0)}_i \,,
\end{equation}
in the Jacobi basis it is convenient to consider explicitly only the coupling to
the ``last'' nucleon (relative to the rest, associated with the Jacobi
momenta $\vec{u}_2$ and $\vec{u}_3$ for three and four nucleons, respectively)
and to include a factor $A$ to account for the remaining nucleons based on the
overall (anti-)symmetry of the wave function.
The relevant charge operator still needs to be expressed in the $3N$/$4N$
partial-wave bases, which is done by first decoupling the spin and angular
degrees of freedom.
The final result involves then the reduced spatial matrix element of the
monopole charge operator, written generically as
$\braket{u;l||\hat{\rho}_{L=0}^{(0)}(\alpha\vec{q})||u';l'}$
for a nucleon pair with relative momentum $u=|\vec{u}|$ and associated
angular quantum momentum $l$.\footnote{Charge operator multipoles $L>0$ do not
contribute to the form factors.}
The detailed expression for this is given in
Ref.~\cite{Konig:2019xxk}.
The factor $\alpha$ arises from expressing the single-particle coordinates in
terms of the Jacobi coordinates and the overall center-of-mass coordinate of the
system.

The spin quantum numbers give rise to the following recoupling factors
for $A=3$:
\begin{multline}
 \braket{u_1 u_2;s|\hat{\rho}^{(0)}_{3,L=0}(\vec{q})|u_1'u_2';s'}
 = ({-}1)^{s_2+J+l_2'} \sqrt{\hat{J}\hat{l_2'}} \\
 \null \times
 \SixJ{l_2}{s_2}{J}{J'}{0}{l_2'}
 \delta_{j_1j_1'} \delta_{l_1l_1'} \delta_{s_1s_1'} \delta_{s_2s_2'}
 \frac{\delta(u_1-u_1')}{u_1^2} \\
 \times \braket{u_2;l_2||\hat{\rho}_{L=0}^{(0)}(\tfrac43\vec{q})||u_2';l_2'} \,.
\end{multline}
with $\hat{l} = (2l+1)$.
Moreover, the various quantum numbers refer to those in Eq.~\eqref{eq:s}, with a
prime added to $s$ indicating primes for all quantum numbers collected within
the state.
For, $A=4$, one obtains, similarly:
\begin{multline}
 \braket{u_1 u_2 u_3;a|\hat{\rho}^{(0)}_{4,L=0}(\vec{q})|u_1'u_2'u_3';a'}
 = ({-}1)^{\frac12 + j_2 + 2j_3 + l_3' + J} \\
 \null \times \sqrt{\hat{j_3}\hat{j_3'}} \sqrt{\hat{J}\hat{l_3'}}
 \SixJ{l_3}{\frac12}{j_3}{j_3'}{0}{l_3'}
 \SixJ{j_3}{j_2}{J}{J'}{0}{j_3'} \\
 \null \times \delta_{j_1j_1'} \delta_{l_1l_1'} \delta_{s_1s_1'}
 \delta_{j_2j_2'} \delta_{l_2l_2'} \delta_{s_2s_2'}
 \frac{\delta(u_1-u_1')}{u_1^2} \frac{\delta(u_2-u_2')}{u_2^2}
 \\
 \times \braket{u_3;l_3||\hat{\rho}_{L=0}^{(0)}(\tfrac32\vec{q})||u_3';l_3'} \,.
\end{multline}

Finally, the isospin matrix elements have to be evaluated for the particular
states of interest.
For \isotope[3]{H}, we have total isospin $T=1/2$ with projection $M_T={-}1/2$.
As we are considering the current coupled to the third nucleon, the relevant
matrix element is
\begin{multline}
 \braket{(t_1\tfrac12)TM_T|\frac{1+\tau_3^{(3)}}{2}|(t_1'\tfrac12)TM_T} \\
 \null = \delta_{t_1t_1'} \sum_{m_1,\mu_3}
 \abs{\CG{t_1}{m_1}{\frac12}{\mu_3}{T}{M_T}}^2
 \frac{1+2\mu_3}{2} \,.
\end{multline}
Similarly, for \isotope[4]{He} (with $T=M_T=0$), the relevant matrix element is
\begin{multline}
 \braket{((t_1\tfrac12)t_2\tfrac12)TM_T|\frac{1+\tau_3^{(3)}}{2}|
 ((t_1'\tfrac12)t_2'\tfrac12)TM_T} \\
 \null = \delta_{t_1t_1'} \delta_{t_2t_2'} \sum_{m_2,\mu_4}
 \abs{\CG{t_2}{m_2}{\frac12}{\mu_4}{T}{M_T}}^2
 \frac{1+2\mu_4}{2} \,.
\end{multline}

\end{document}